\documentclass{article}

\usepackage{PRIMEarxiv}

\usepackage[utf8]{inputenc} 
\usepackage[T1]{fontenc}    
\usepackage{hyperref}       
\usepackage{url}            
\usepackage{booktabs}       
\usepackage{amsfonts}       
\usepackage{nicefrac}       
\usepackage{microtype}      
\usepackage{lipsum}
\usepackage{fancyhdr}       
\usepackage{graphicx}       
\usepackage{amsmath}
\graphicspath{{media/}}     
\usepackage{caption}
\title{IR Models and the COVID-19 Pandemic: A Comparative Study of Performance and Challenges}

\author{
  Moksh Shukla \\ 
  Indian Institute of Technology Kanpur \\
  Kanpur, India\\
  \texttt{moksh@iitk.ac.in} \\
   \And
  Nitik Jain \\
 Indian Institute of Technology Kanpur \\
  Kanpur, India\\
  \texttt{nitik@iitk.ac.in} \\
  \And
  Shubham Gupta \\
  Indian Institute of Technology Kanpur \\
  Kanpur, India \\
  \texttt{shubgupt@iitk.ac.in} \\
}

\begin{document}
\maketitle

\begin{abstract}
This research study investigates the efficiency of different information retrieval (IR) systems in accessing relevant information from the scientific literature during the COVID-19 pandemic. The study applies the TREC framework to the COVID-19 Open Research Dataset (CORD-19) and evaluates BM25, Contriever, and Bag of Embeddings IR frameworks. The objective is to build a test collection for search engines that tackle the complex information landscape during a pandemic. The study uses the CORD-19 dataset to train and evaluate the IR models and compares the results to those manually labeled in the TREC-COVID IR Challenge. The results indicate that advanced IR models like BERT and Contriever better retrieve relevant information during a pandemic. However, the study also highlights the challenges in processing large datasets and the need for strategies to focus on abstracts or summaries. Overall, the research highlights the importance of effectively tailored IR systems in dealing with information overload during crises like COVID-19 and can guide future research and development in this field.
\end{abstract}

\keywords{ Information Retrieval Systems\and COVID-19 \and TREC framework }

\section{Introduction}

Accessing the scientific literature through the most effective information retrieval (IR or search) technologies is crucial for locating evidence. Hence, rapid implementation of IR systems optimised for such a context and a comparison of their efficacy were required. In consequence, the COVID-19 global health crisis has exacerbated the issue. The immediate need for information on a worldwide scale has resulted in an exponential increase in scientific literature publication. Relevant and dependable information has become an urgent necessity.

Concerning the application of IR in such a pandemic, there are numerous fundamental research concerns that must be answered, including the identification of essential IR modalities, the development of domain-specific search engines, and the quantitative evaluation of search engine performance. In an effort to combat the pandemic, massive COVID-19-related corpora are being compiled, often containing erroneous data that is no longer amenable to human analysis.

The challenge evaluation is a standard method for evaluating IR systems on a wide scale. The Text Retrieval Conference (TREC), organised by the US National Institute of Standards and Technology, is the largest and most well-known approach (NIST). The TREC framework was applied to the COVID-19 Open Research Dataset (CORD-19), a dynamic repository of scientific papers on COVID-19 and historical coronavirus research connected to COVID-19. The dataset is described in full under Section corpus. The major objective of the TREC-COVID Competition was to develop a collection of test data for evaluating search engines' ability to navigate the complicated information landscape during events such as a pandemic. In this research, we evaluate various Information Retrieval frameworks, including BM25, Contriever, Bag of Embeddings, etc., based on their ability to rank documents according to their relevance to input queries.

In this study, we begin by utilising document data from the CORD-19 dataset and pre-processing it with its metadata. We opted to examine several IR models because the hand classified data categorising documents as relevant, slightly relevant, or irrelevant was unavailable. Our baseline model is BM25, which compares the relevance of searches to the document by utilising the Abstract and Title. Additionally, we retrieve the list of relevant documents from other faster models and compare them to the standard documents. Finally, we compare our results to those manually labelled according to the TREC-COVID IR Challenge. 

\section{Related Work}
The COVID-19 pandemic has led to a global surge in scientific literature, and the need for efficient information retrieval (IR) systems has become more critical than ever. The CORD-19 dataset, a large-scale open access resource of scientific papers on COVID-19 and related historical coronavirus research, has emerged as a valuable resource for COVID-19 research. As a result, several studies have been conducted to explore the use of IR systems and natural language processing (NLP) and machine learning (ML) techniques to improve IR performance on the CORD-19 dataset.

One notable study conducted by \cite{wang2021covid} developed the CORD-19 dataset, which has been widely used in COVID-19 research. Other studies have used NLP and ML techniques to improve IR performance on the CORD-19 dataset. For example, \cite{rizk2021deep} developed a deep learning-based text mining framework for COVID-19 literature, while \cite{si2021overview} used unsupervised topic modeling and machine learning techniques to analyze the COVID-19 "infodemic" using the CORD-19 dataset.

In addition to improving IR performance, the CORD-19 dataset has been used for drug repurposing and knowledge graph construction. \cite{wang2021covid} developed a COVID-19 literature knowledge graph construction and drug repurposing report generation system using the CORD-19 dataset. \cite{khare2021mining} used the CORD-19 dataset to mine the literature for COVID-19 research and identify potential drug candidates for further investigation.

Studies have also evaluated the CORD-19 dataset and its use in IR systems. \cite{tariq2020cord} provided an overview of the CORD-19 dataset and its potential uses, while \cite{si2021overview} applied domain adaptation techniques to improve IR performance on the CORD-19 dataset. Overall, the CORD-19 dataset has been widely used and evaluated in the context of COVID-19 research, particularly in the development and evaluation of IR systems using NLP and ML techniques, as well as for drug repurposing and knowledge graph construction.

The COVID-19 pandemic has highlighted the need for rapid implementation and evaluation of IR systems. The CORD-19 dataset has emerged as a valuable resource for COVID-19 research and has been extensively studied to improve IR performance, analyze the "infodemic," and identify potential drug candidates. As such, the CORD-19 dataset has become a crucial component of COVID-19 research, with many studies evaluating its effectiveness and exploring its potential uses.

In conclusion, the CORD-19 dataset has been widely used and evaluated in the context of COVID-19 research, particularly in the development and evaluation of IR systems using NLP and ML techniques, as well as for drug repurposing and knowledge graph construction. The dataset has provided valuable insights and has played a significant role in advancing COVID-19 research.

\section{Document Set Description}\label{corpus}
TREC-COVID makes use of the CORD-19 document collection. CORD-19 contains fresh articles and preprints on COVID-19, as well as previous studies on coronaviruses such as SARS and MERS. The CORD-19 release on April 10, 2020, which will be utilised for the first round of TREC-COVID, has 51K papers, with full text available for 39K. We train and evaluate our models using whole \textbf{51000} documents.

\subsection{File Structure:}
\begin{enumerate}
    \item \textbf{\texttt{CORD-19}} - folder holding the CORD-19 papers and metadata as of May 19, 2020. 
 \item \textbf{\texttt{topics-rnd3.csv}} - file containing each topic's \textit{topic-id, query, question}, and \textit{narrative}.
 \item \textbf{\texttt{docids-rnd3.txt}} - the set of documents that can be projected to be relevant; these papers do not include those that have been rated in prior rounds for a topic.
 \item \textbf{\texttt{qrels.csv}} - human annotated relevant and irrelevant document categorization for the 40 queries.
 \end{enumerate}

 \subsection{Topics:}
 
TREC-COVID topics were written by its organisers with biomedical training, and were inspired by consumer questions submitted to the National Library of Medicine, discussions by medical influencers on social media, and suggestions solicited on Twitter in late March 2020 via the \#COVIDSearch tag. They are representative of the pandemic's high-level concerns. An initial set of 30 subjects was established, with 5 new topics added for each subsequent round. As a result, we ran our retrieval on \textbf{40} topics, which can be accessed in the \textbf{\texttt{topics-rnd3.csv}} file.

The subject file is an xml file that contains all of the round's topics. A topic is formatted as follows (this is an example topic, not part of the official topic set):

\begin{figure}[h!]
    \centering
    \includegraphics[width=\textwidth]{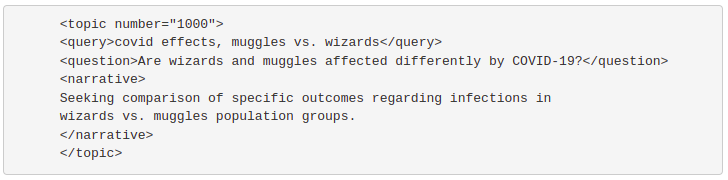}
    \label{fig:example-topic}
\end{figure}

Each topic is composed of three fields:
\begin{enumerate}
    \item \texttt{query:} a short keyword query
    \item \texttt{question:} a more precise natural language question
    \item \texttt{narrative:} a more detailed description that expands on the question, frequently specifying specific types of facts that would fall under the topic score
\item \texttt{average:} in addition, we generate another set of questions that are the average of the aforementioned three embeddings. This is a frequently used and acknowledged technique in NLP.
\end{enumerate}

\begin{table}[h!]
\centering
\begin{tabular}{|c|c|c|}
\hline
\textbf{Query} & \textbf{Question} & \textbf{Narrative} \\
\hline
\begin{tabular}{@{}c@{}}Coronavirus \\ response to \\ weather changes\end{tabular} & \begin{tabular}{@{}c@{}}How does the coronavirus \\ respond to changes in the \\ weather?\end{tabular} & \begin{tabular}{@{}c@{}} Seeking a variety of information about virus survival in \\ various weather/climate settings, as well as \\ information about virus transmission in \\ various climatic circumstances.\end{tabular} \\
\hline
\begin{tabular}{@{}c@{}}Coronavirus \\ social distancing \\ impact\end{tabular} & \begin{tabular}{@{}c@{}}Has social distancing had \\ an impact on slowing the \\ spread of COVID-19?\end{tabular} & \begin{tabular}{@{}c@{}} Seeking particular information on studies that have \\ investigated COVID-19 transmission in one or more \\ social (or non-social) \\ techniques.\end{tabular} \\
\hline
\begin{tabular}{@{}c@{}}Coronavirus \\ outside body\end{tabular} & \begin{tabular}{@{}c@{}}How long can the \\ coronavirus live outside the \\ body?\end{tabular} & \begin{tabular}{@{}c@{}} Seeking a variety of information about the virus's survivability \\ outside the human body (surfaces, liquids, etc.) \\ while remaining alive for \\ transmission to another human.
\end{tabular} \\
\hline
\begin{tabular}{@{}c@{}}coronavirus \\ asymptomatic\end{tabular} & \begin{tabular}{@{}c@{}}What is known about those \\ infected with Covid-19 but \\ are asymptomatic?\end{tabular} & \begin{tabular}{@{}c@{}}Studies on patients who are known to be Covid-19 infected \\ but show no symptoms?\end{tabular} \\
\hline
\begin{tabular}{@{}c@{}}Coronavirus \\ hydroxy- \\ chloroquine\end{tabular} & \begin{tabular}{@{}c@{}}What evidence is there for \\ the value of \\ hydroxychloroquine in \\ treating Covid-19?\end{tabular} & \begin{tabular}{@{}c@{}}
Basic scientific or clinical research evaluating the \\ benefits and risks of treating Covid-19 with \\ hydroxychloroquine.
\end{tabular} \\
\hline
\end{tabular}
\caption{Few illustrative examples of topics for TREC-COVID task.}
\label{table:example_queries}
\end{table}


\section{Methodology}
\subsection{Pre-processing}
In the data-set, for each document, we are provided with two types of JSON files: \texttt{pdf\_json} and \texttt{pmc\_json}, which differ only in their format. For simplicity, we used only the \texttt{pdf\_json} form. We first cleared the data set from NaN values. Then we extracted the abstract of each document and stored it separately along with the document ID. Later, we used this data for various tasks detailed in his report.

For the BM25 model as well as the Bag of Embeddings model the pre-processing done was very similar to the one we did in our class assignments. The corpus consisted of .json parsed files for each research article as mentioned above and we leveraged that to read the abstract, title and the body text as well (only for BM25), created a single string for the whole article and moved ahead with the pre-processing and cleaning part.
The pre-processing and the cleaning involved the following things:-
\begin{itemize}
\item {Lower Casing}
\item {Email and url removal}
\item {Non ASCII removal}
\item {Special Characters removal}
\item {Stop words removal}
\item {Stripping extra white space and stemming}
\item {Tokenizing}
\end{itemize}

For the queries also same pre-processing was done before running the query
\subsection{Models}

We implemented and compared the following IR models:

\subsubsection{BM25 - The Baseline Model \cite{bm25}}
A type of ranking function that ranks a set of documents based on the query terms that exist in each document, independent of how the query terms inside a document are related. The BM25 term weighting algorithms have been widely and successfully employed in a variety of collections and search activities. 
\\

\textbf{The ranking function:} BM25 is a bag-of-words retrieval algorithm that ranks a set of documents based on the query phrases that exist in each document, regardless of how close they are to each other. It refers to a group of scoring functions that have somewhat varied components and parameters. The following is one of the more notable instantiations of the function.

Given a query $Q$, containing keywords $q_{1}, \ldots, q_{n}$, the BM25 score of a document $D$ is:

$$
\operatorname{score}(D, Q)=\sum_{i=1}^{n} \operatorname{DF}\left(q_{i}\right) \cdot \frac{f\left(q_{i}, D\right) \cdot\left(k_{1}+1\right)}{f\left(q_{i}, D\right)+k_{1} \cdot\left(1-b+b \cdot \frac{|D|}{\mathrm{svgdl}}\right)}
$$

where $f\left(q_{i}, D\right)$ is $q_{i}$ 's term frequency in the document $D,|D|$ is the length of the document $D$ in words, and avgdl is the average document length in the text collection from which documents are drawn. $k_{1}$ and $b$ are free parameters, usually chosen, in absence of an advanced optimization, as $k_{1} \in[1.2,2.0]$ and $b=0.75 \mathrm{IDF}\left(q_{i}\right)$ is the IDF (inverse document frequency) weight of the query term $q_{i}$. It is usually computed as:

$$
\operatorname{IDF}\left(q_{i}\right)=\ln \left(\frac{N-n\left(q_{i}\right)+0.5}{n\left(q_{i}\right)+0.5}+1\right)
$$
where $N$ is the total number of documents in the collection, and $n\left(q_{i}\right)$ is the number of documents containing $q_{i}$. 

\subsubsection{Contriever \cite{contriever}}
Neural network-based information retrieval has achieved state-of-the-art performance on datasets and problems where extensive training sets are available. Unfortunately, they do not translate well to new domains or applications without training data and are frequently surpassed by unsupervised term-frequency approaches like BM25. Contriever is a basic self-supervised IR model based on contrastive learning that is competitive with BM25.
\\

\textbf{Contrastive Learning:}
Contrastive learning is a method that takes advantage of the fact that every document is distinctive in some way. This signal is the only information provided in the absence of manual monitoring. Using a contrastive loss (citecontrastive), the final algorithm learns by differentiating between documents. This loss compares pairs of document representations that are either positive (from the same document) or negative (from distinct documents). Formally, the contrastive textttInfoNCE loss is defined as follows: 
$$
\mathcal{L}\left(q, k_{+}\right)=\frac{\exp \left(s\left(q, k_{+}\right) / \tau\right)}{\sum_{i=0}^{K} \exp \left(s\left(q, k_{i}\right) / \tau\right)},
$$
This loss increases the relevance value of comparable examples and decreases the relevance score of dissimilar ones. This loss function can also be interpreted as follows: given the query representation $q$, the objective is to recover or retrieve the representation $k_{+}$ corresponding to the positive document from among all the negatives $k_i$. The left-hand side representations in the score $s$ are referred to as questions, whereas the right-hand side representations are referred to as keys.
\\

\textbf{Building positive pairings from a single document:} A fundamental feature of contrastive learning is how to construct positive pairs from a single input. This stage in computer vision is applying two independent data augmentations to the same image, yielding two "views" that constitute a positive pair. While contriever primarily considers similar independent text modifications, they also investigate dependent transformations that aim to lessen the correlation across perspectives. They are: 
\begin{enumerate}
    \item Inverse Cloze Task
    \item Independent cropping
    \item Further data augmentation.
\end{enumerate}

\textbf{Building large sets of negative pairs:} The maintenance of a high number of negative pairs is a crucial feature of contrastive learning. The majority of common frameworks handle negatives differently, and the two utilised here are:
\begin{enumerate}
    \item Negative pairs within a batch.
    \item Negative pairs across a batch.
\end{enumerate}

\begin{figure}[!h]
    \centering
    \includegraphics[width=\textwidth]{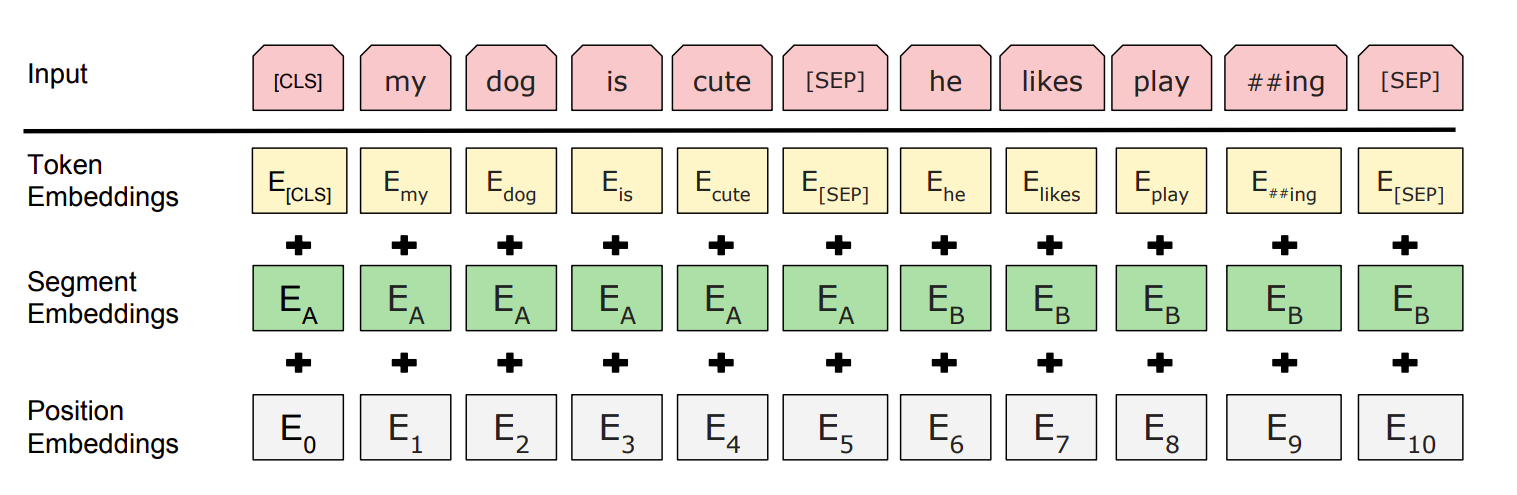}
    \caption{BERT input representation. The input embeddings are the sum of the token embeddings, the segmentation embeddings and the position embeddings}
    \label{fig:bert}
\end{figure}

\subsubsection{Bag-of-Embeddings \cite{boe}}
Instead of a single point in the semantic space, a word is represented as a probability distribution that encompasses the complete semantic area. It corresponds specifically to the multivariate normal distribution with spherical covariance. Given a second embedded word, the "overlap" between these two distributions can be used to determine the degree of similarity between the two words.

\subsubsection{BERT Embeddings \cite{bert}}    
Google created the state-of-the-art Bidirectional Encoder Representation from Transformers (BERT) technique for natural language processing pre-training. Our corpus will be embedded with a BERT embedding that has been pre-trained.


\section{Results}




In this section of our study paper, we compare the retrieval performance of three distinct models, namely BERT, Contriever, and BM25. To evaluate the models, the top 50 most relevant documents were extracted for each of the 40 questions in the dataset using each of the three models. This procedure generated three distinct lists of document IDs for each query, one list for each model.

To establish a fair comparison across the models, we determined the average score for the extracted documents, as described in the section on methodology. We next did a pairwise intersection of the document ID lists from the three models for each query. This analysis generated three unique pairs: BERT and BM25, BERT and Contriever, and BM25 and Contriever.

The amount of shared documents obtained for each query by each pair of models provides insight into the degree of overlap between the pages deemed relevant by each pair. Finally, we computed the mean and standard deviation of the number of common documents across all queries for each pair in order to standardise the results and gain a deeper understanding of the consistency and agreement across the models in terms of document retrieval performance.

The following table summarizes our findings:

\begin{table}[h!]
\centering
\begin{tabular}{|c|c|c|c|}
\hline
\textbf{Model Comparison} & \textbf{BERT-Contriever} & \textbf{BERT-BM25} & \textbf{BM25-Contriever} \\
\hline
Mean & 36.725 & 18.1 & 11.775 \\
\hline
Standard Deviation & 17.927 & 13.647 & 11.014 \\
\hline
\end{tabular}
\caption{Summary of the results comparing the retrieval performance of BERT, Contriever, and BM25}
\label{table:results}
\end{table}


The mean and standard deviation are greatest for the BERT-Contriever pair, followed by the BERT-BM25 pair, and then the BM25-Contriever pair. This indicates that BERT and Contriever have the greatest degree of agreement about document retrieval, but BM25 has the least overlap with both BERT and Contriever. By analysing the mean and standard deviation of the number of common documents obtained by each pair of models, we can draw inferences regarding the similarity and dissimilarity of their performance in obtaining relevant documents in our research topic.

\section{Conclusion}
In conclusion, our study emphasises the significance of information retrieval systems for conveniently gaining access to essential scientific publications during a pandemic. The application of the TREC framework to the COVID-19 Open Research Dataset (CORD-19) allowed for the evaluation and comparison of various information retrieval (IR) frameworks such as BM25, Contriever, and Bag of Embeddings, with the primary objective of constructing a test collection for search engines dealing with the complex information landscape during events such as a pandemic.

The CORD-19 dataset, consisting of publications and preprints on COVID-19 and relevant historical coronavirus research, was utilised to train and assess the IR models. Our methodology entailed preprocessing document data, utilising metadata and abstracts to compare the performance of several IR models, with BM25 serving as the baseline. In addition, the outcomes were compared to those manually labelled in the TREC-COVID IR Challenge.


According to the results of our investigation, the performance of the IR models varied, with BERT and Contriever exhibiting the most document overlap, followed by BERT-BM25 and BM25-Contriever. These findings demonstrate the capacity of advanced IR models such as BERT and Contriever to retrieve pertinent information during a pandemic. However, the study also revealed the difficulties associated with processing huge datasets due to limited computer resources, necessitating the employment of tactics such as concentrating on abstracts and summaries.
This study highlights the significance of well-tailored IR systems for managing the information deluge during crises such as the COVID-19 pandemic. This study's findings can influence future research and development of IR systems in order to enhance their performance in quickly developing situations, hence facilitating decision-making and response efforts.

\section{Challenges and Future Work}
Our main challenge was the availability of adequate computational resources to process the vast dataset. Despite our best efforts, including taking only the abstract and summary, using a summarised version, or limiting the number of documents processed, we were constrained by the limited computational resources.
Moreover, we encountered another challenge when utilizing Sentence-BERT on an entire paragraph. To address this issue, we initially employed Pegasus, followed by S-BERT.

In future work, we aim to explore more advanced algorithms and hybrid models to enhance retrieval performance. Our current approach relies on vanilla models for all three tasks. Additionally, fine-tuning BERT specifically for the COVID-19 dataset represents another promising avenue for future research.

\bibliographystyle{unsrt}  
\bibliography{references}

\end{document}